\documentclass[10pt,letterpaper]{article}
\usepackage{opex3}
\usepackage{color}
\usepackage{amsmath}
\usepackage{cite}
\begin{document}

\title{Dynamics of mode-coupling-induced microresonator frequency combs in normal dispersion}

\author{Jae~K.~Jang$^{1,*}$, Yoshitomo~Okawachi$^1$, Mengjie~Yu$^{1,2}$, Kevin~Luke$^2$, Xingchen~Ji$^{2,3}$, Michal~Lipson$^3$, and Alexander~L.~Gaeta$^1$}

\address{$^1$Department of Applied Physics and Applied Mathematics, Columbia University, New York, NY 10027\\
$^2$School of Electrical and Computer Engineering, Cornell University, Ithaca, NY 14853\\
$^3$Department of Electrical Engineering, Columbia University, New York, NY 10027}

\email{$^*$jj2837@columbia.edu} 



\begin{abstract}
We experimentally and theoretically investigate the dynamics of microresonator-based frequency comb generation assisted by mode coupling in the normal group-velocity dispersion~(GVD) regime. We show that mode coupling can initiate intracavity modulation instability~(MI) by directly perturbing the pump-resonance mode. We also observe the formation of a low-noise comb as the pump frequency is tuned further into resonance from the MI point. We determine the  phase-matching conditions that accurately predict all the essential features of the MI and comb spectra, and extend the existing analogy between mode coupling and high-order dispersion to the normal GVD regime. We discuss the applicability of our analysis to the possibility of broadband comb generation in the normal GVD regime.
\end{abstract}



\section{Introduction}

Optical frequency comb generation based on nonlinear microresonators~\cite{delhaye_optical_07,savchenkov_tunable_08,agha_theoretical_09,levy_cmos_10,razzari_cmos_10,kippenberg_microresonator_11, foster_silicon_11,ferdous_spectral_11,savchenkov_kerr_12,saha_modelocking_13,jung_optical_13,herr_temporal_14,hausmann_diamond_14,liang_generation_14, liu_investigation_14,griffith_silicon_15,xue_normal_15,xue_modelocked_15,huang_low_15,yi_soliton_15,webb_measurement_16,brasch_photonic_16, joshi_thermally_16,wang_intracavity_16} is a rapidly evolving field of research that could potentially benefit such diverse fields as optical metrology, spectroscopy, optical waveform synthesis and telecommunications~\cite{kippenberg_microresonator_11}. Not only are such combs attractive for applications, but they also exhibit rich nonlinear dynamics that are of fundamental interest such as modelocking and cavity soliton formation~\cite{saha_modelocking_13,herr_temporal_14,yi_soliton_15,brasch_photonic_16,joshi_thermally_16,wang_intracavity_16}, and dispersive wave emission~\cite{brasch_photonic_16}, which were all observed previously in driven passive fiber cavities~\cite{leo_temporal_10,jang_observation_14}.

Many studies have been dedicated to understanding the mechanisms that drive comb generation in microresonators~\cite{agha_theoretical_09,kippenberg_microresonator_11,chembo_modal_10,hansson_dynamics_13}. It is now well understood that microresonator comb generation relies on the Kerr nonlinearity and the associated multimode four-wave-mixing~(FWM). Intracavity modulation instability~(MI)~\cite{lugiato_spatial_87,haelterman_dissipative_92,coen_modulational_97,hansson_dynamics_13} is specifically cited as a key nonlinear process that governs the initial dynamics of the comb growth. Anomalous group-velocity dispersion~(GVD) inherently allows phase-matching of the MI process, and for this reason, microresonator combs have been investigated primarily in resonators with anomalous GVD. More recently, various studies have reported evidence of frequency combs in resonators characterized by normal GVD~\cite{savchenkov_kerr_12,liu_investigation_14,liang_generation_14,xue_modelocked_15,xue_normal_15}. Intracavity MI is possible in the normal GVD regime due to the cavity boundary condition which provides an extra degree of freedom, via the cavity detuning, to the MI phase-matching condition~\cite{haelterman_dissipative_92,coen_modulational_97}. In the microresonator geometry, on the other hand, intracavity MI process is difficult to achieve reliably in the normal GVD regime since it occurs in the red-detuned region of the Kerr-bistability cycle~\cite{xue_normal_15} where the resonators are thermally unstable~\cite{carmon_dynamical_04}. An alternative route to normal GVD comb generation can be realized by exploiting coupling between different mode families that are supported simultaneously by the resonators. Such mode coupling occurs when longitudinal resonances of different spatial~\cite{carmon_static_08} or polarization~\cite{ramelow_strong_13} mode families overlap in frequency, causing shifting of the coupled resonances and locally altering the GVD.

The effects of mode coupling in microresonator comb generation in the anomalous GVD regime have been shown to be mostly detrimental and can inhibit soliton generation in the most severe cases~\cite{herr_mode_14}. In addition, it has been suggested that in this regime, mode coupling can play a role similar to high-order dispersion, in which case comb spectra display features similar to those of dispersive waves~\cite{zhou_stability_15,webb_measurement_16,matsko_optical_16,yang_spatial_16}. In contrast, mode coupling has been reported to aid comb generation in normal GVD microresonators~\cite{savchenkov_kerr_12,liu_investigation_14,liang_generation_14,xue_modelocked_15,xue_normal_15} which in some cases corresponds to the formation of dark pulse-like structures in the time domain~\cite{xue_modelocked_15}. This latter work also hinted that mode coupling may be responsible for generating the initial MI sidebands in which phase-matching was achieved by shifting one of the sideband modes far away from the pump-resonance mode while allowing the pump to operate in the blue-detuned regime of the Kerr bistability and remain thermally stable~\cite{carmon_dynamical_04}. Nevertheless, in comparison to the dynamics of the anomalous GVD regime, the comb generation in the normal GVD regime is relatively unexplored.

In this Article, we investigate in detail the dynamical behavior of comb generation triggered by mode coupling in a microresonator characterized by normal GVD. In contrast to~\cite{xue_modelocked_15} where mode coupling influenced one of the sideband modes far away from the pump, we focus on the scenario in which the pumped longitudinal resonance mode is directly perturbed by mode coupling. We show that the initial comb spectrum consists of a pair of multiple-mode-spaced sidebands exhibiting low amplitude noise, which is reminiscent of intracavity MI. The mode coupling in this case shifts and aligns the pumped resonance mode with the pump laser wavelength, leading to strong build-up of the intracavity power and allowing the associated parametric gain to overcome the cavity roundtrip loss. We subsequently tune the pump frequency further into resonance and study the generated low-noise comb. This comb exhibits spectral features similar to dispersive waves, and we explain them in terms of the analogy between high-order dispersion and mode coupling. Such analogy was already applied in the anomalous GVD regime~\cite{zhou_stability_15,webb_measurement_16,matsko_optical_16,yang_spatial_16}. Here we effectively extend this concept to the normal GVD case and show that our study is generally applicable to normal GVD combs formed under similar conditions. The potential of broadband comb generation is also discussed in this context.

\section{Numerical and experimental characterization of mode coupling}

In our experiments, we employ an oxide-clad silicon nitride~($\mathrm{Si}_3\mathrm{N}_4$) microresonator with a free-spectral-range~(FSR) of 200~GHz. The cross-section of the $\mathrm{Si}_3\mathrm{N}_4$ waveguide is $730$~nm high and $3000$~nm wide, which ensures that the fundamental TE mode of the waveguide is in the normal GVD regime near the pump wavelength. The resonator is pumped with a tunable continuous-wave~(CW) laser which is tuned from $1541$ to $1548$~nm depending on the experiment. The typical pump power in the integrated bus waveguide, which is evanescently coupled to the resonator, varies from $5$ to $10$~mW. Comb generation is initiated when the frequency of the CW laser is externally tuned into resonance from the blue-detuned regime, and no active stabilization of the pump nor cavity was utilized. The output is collimated and collected for analysis by an optical spectrum analyzer, an RF spectrum analyzer, and an oscilloscope.

\begin{figure}[t]
\centerline{\includegraphics[width=0.8\columnwidth,clip]{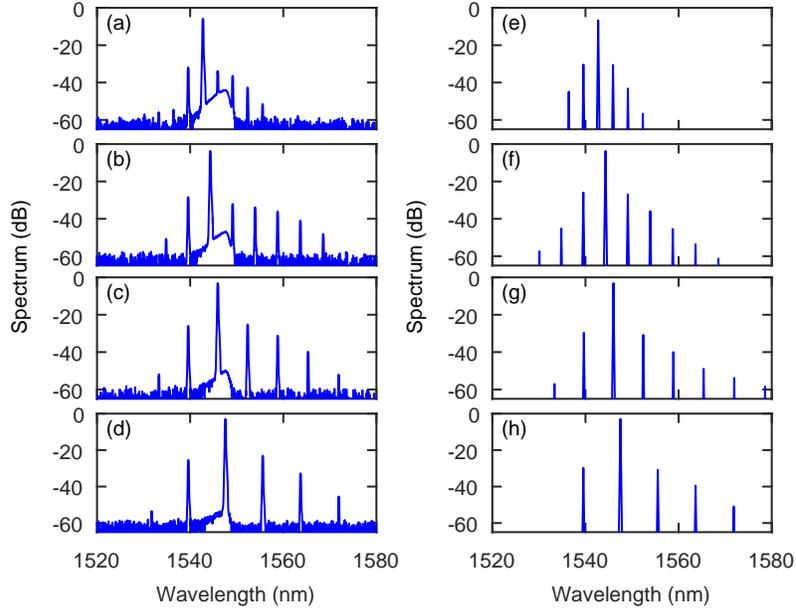}}
\caption{The experimental [(a)--(d)] and simulated [(e)--(h)] optical spectra as the pump wavelength is tuned over 4 consecutive resonances. The first blue-detuned sideband relative to the pump remains pinned at $\sim$ 1540~nm.}
\label{fig:fig1}
\end{figure}

Due to its relatively large width, this $\mathrm{Si}_3\mathrm{N}_4$ waveguide supports multiple spatial mode families. Each mode family exhibits a unique mode spacing and as a consequence, it is possible for longitudinal resonances belonging to two distinct spatial mode families to overlap in frequency. Waveguide imperfections in the form of surface roughness could lead to coupling between the overlapping resonances and to breaking of the frequency degeneracy, that is, local shifting and splitting of the resonance modes~\cite{carmon_static_08,ramelow_strong_13}. This effect results in significant local disruptions in the GVDs, experienced by both the mode families involved. It was reported that such a GVD modification can facilitate the phase-matched FWM process, which can initiate comb generation in normal GVD resonators~\cite{savchenkov_kerr_12,xue_modelocked_15}. In particular, a recent study has shown that the vicinity of mode coupling can act as a `pinning' site where one of the nearest generated sidebands is robustly anchored over a significant tuning range of the pump wavelength~\cite{liu_investigation_14}. We verify this result for our resonator by tuning the pump laser wavelength within proximity of the mode-coupling region. The results are summarized in Figs.~\ref{fig:fig1}(a)--(d). It can be seen that although the laser is tuned over 4 resonances from $1542.8$~nm to $1547.6$~nm, the first blue-detuned sideband (relative to the pump) remains anchored at approximately $1540$~nm, implying the presence of mode coupling.

Direct experimental characterization of the resonator GVD and mode coupling proved to be difficult due to the presence of a large number of resonances belonging to different spatial mode families. Instead, we estimate the coupling strength by performing numerical simulations based on a generalized Lugiato-Lefever equation~(LLE)~\cite{lugiato_spatial_87,haelterman_dissipative_92,coen_modeling_12,chembo_spatiotemporal_13,lamont_route_13}
\begin{equation}\label{eq:LLE}
t_\mathrm{R}\frac{\partial E(t,\tau)}{\partial t} = \bigg[-\alpha-i\delta_0+iL\sum_{m \geq 2} \frac{\beta_m}{m!} \bigg(i\frac{\partial}{\partial\tau} \bigg)^m+i\gamma L|E|^2 \bigg]E(t,\tau) + \sqrt{\theta}E_\mathrm{in},
\end{equation}
where $E(t,\tau)$ is the intracavity field envelope, $t$ is the `slow' evolution time on the order of the cavity roundtrip time $t_\mathrm{R}$ while $\tau$ is a `fast' time axis that travels at the group velocity of light around the cavity, $L$ is the cavity length, $\alpha$ describes the total cavity roundtrip loss, $\gamma$ is the Kerr-nonlinearity coefficient, and $\beta_m$ is the $m^\mathrm{th}$ order dispersion coefficient. Finally, $\theta$ is the input coupling coefficient, $E_\mathrm{in}$ is the pump field amplitude such that $|E_\mathrm{in}|^2$ is the pump power measured in W, and $\delta_0$ is the pump phase detuning with respect to the nearest resonance. We use the following parameters to match our experiment: $t_\mathrm{R} = 1/\mathrm{FSR} = 5~\mathrm{ps}$, $L = 810~\mu\mathrm{m}$, $\alpha = 7.6\times 10^{-4}$, $\gamma = 0.63~\mathrm{W^{-1}m^{-1}}$, $\beta_2 = 36~\mathrm{ps^2/km}$, $\beta_3 = 0.016~\mathrm{ps^3/km}$, $\beta_4 = 9.0\times 10^{-4}~\mathrm{ps^4/km}$, and $\theta = 3.5\times 10^{-4}$. Although high-order dispersion (third-order and higher) plays a negligible role here, we have included it for completeness. The effect of mode coupling is incorporated phenomenologically in our simulation by using the empirical two-parameters model~\cite{herr_mode_14}. For the fundamental TE mode, we estimate the mode-coupling-induced resonance shift to be $0.064$\% of the FSR ($\simeq 130$~MHz) at the position of mode coupling at $1541.2$~nm. As can be seen in Figs.~\ref{fig:fig1}(e)--(h), our simulation results based on these parameters are in good qualitative agreement with the experimental results.

\section{Mode-coupling-assisted intracavity modulation instability}

\begin{figure}[b]
\centerline{\includegraphics[width=0.8\columnwidth,clip]{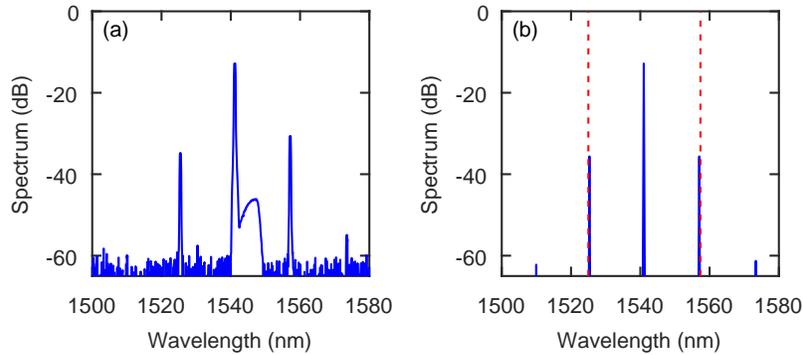}}
\caption{(a) The experimental and (b) numerical spectra of intracavity modulation instability. The red dashed lines indicate the predicted positions of the first sideband pair based on Eq.~(\ref{eq:MIPM}).}
\label{fig:fig2}
\end{figure}

We examine the comb generation dynamics in the presence of mode coupling. As the pump laser is tuned into resonance at 1541.2~nm, we initially observe a pair of sidebands separated by multiple-FSR's [Fig.~\ref{fig:fig2}(a)]. The corresponding simulated spectrum [Fig.~\ref{fig:fig2}(b)] is in good agreement with the experiment and is similar to primary combs that are routinely reported in anomalous GVD resonators as the initial stage of comb generation driven by intracavity MI~\cite{haelterman_dissipative_92,coen_modulational_97}. In this light, we seek an analogous explanation in the normal GVD case. The phase-matching condition of intracavity MI can be expressed as~\cite{haelterman_dissipative_92,coen_modulational_97}
\begin{equation}\label{eq:MIPM}
L \sum_{m \geq 2} \frac{\beta_m}{m!}\Omega^m + 2\gamma L|E_0(\delta_0')|^2 - \delta_0 = 0,
\end{equation}
where $\Omega$ is the frequency detuning of the phase-matched sideband relative to the pump, and $E_0(\delta_0')$ is the steady-state CW intracavity field of the pump mode. For the pump mode, the shifted detuning $\delta_0' = \delta_0 - \Delta\omega t_{\mathrm{R}}$, where $\Delta\omega$ is the mode-coupling-induced resonance frequency shift, is the relevant detuning. Accordingly in the phase-matching condition Eq.~(\ref{eq:MIPM}), we use the modified intracavity power $|E_0(\delta_0')|^2$, where $E_0(\delta_0')$ is computed by calculating the steady-state CW ($\partial/\partial t = 0 = \partial/\partial\tau$) field of Eq.~(\ref{eq:LLE}) with $\delta_0'$ replacing $\delta_0$. Sideband modes are not affected by mode coupling and thus, the unshifted detuning $\delta_0$ is used in the third term in Eq.~(\ref{eq:MIPM}). We apply this modified MI phase-matching condition to the simulated plot in Fig.~\ref{fig:fig2}(b) and obtain very accurate predictions of the phase-matched frequencies where the first sideband pair is expected, as displayed by two vertical red dashed lines. In this sense, this process can be interpreted as mode-coupling-assisted intracavity MI in a normal GVD resonator.

Intracavity MI can occur only if two general requirements, namely phase-matching and power threshold $P_{\mathrm{th}} = \alpha/(\gamma L)$, are simultaneously satisfied~\cite{haelterman_dissipative_92,coen_modulational_97}. For normal GVD resonators in the absence of mode coupling, MI can only occur over a small parameter region within the effectively red-detuned regime of the Kerr-bistability cycle where thermal soft-locking is not possible~\cite{carmon_dynamical_04}. Analysis based on our numerical parameters reveals that our system operates outside the bistability cycle. In this case, the pumped mode is directly perturbed by mode coupling such that the steady-state intracavity power is boosted above the threshold power at which the parametric gain exceeds the cavity roundtrip loss. This clearly distinguishes our current work from~\cite{xue_modelocked_15} in which mode coupling occurred far from the pumped resonance. Indeed, the significance of mode coupling in our system is confirmed by the fact that in its absence, the intracavity power stabilizes at a level far below the threshold and no oscillation occurs.

\begin{figure}[b]
\centerline{\includegraphics[width=0.8\columnwidth,clip]{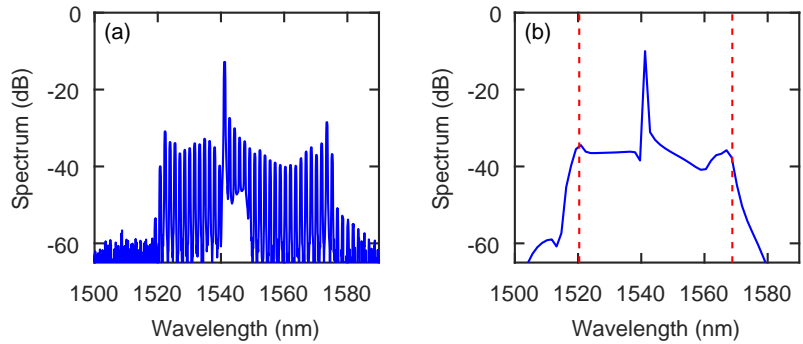}}
\caption{(a) The experimental and (b) numerical spectra of the comb generated via mode coupling. The red dashed lines show theoretical predictions of the secondary spectral peaks.}
\label{fig:fig3}
\end{figure}

\begin{figure}[t]
\centerline{\includegraphics[width=0.6\columnwidth,clip]{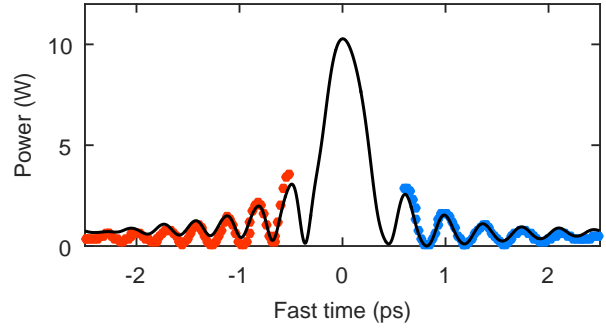}}
\caption{The numerical temporal power profile (black) corresponding to Fig.~\ref{fig:fig3}(b). The red and blue circles display the theoretical asymptotic traces of the decaying oscillatory tails.}
\label{fig:fig4}
\end{figure}

\section{Analogy between high-order dispersion and mode coupling in normal GVD}

As the pump laser is tuned further into resonance, the spectral gaps fill in with additional comb lines to form a 1-FSR-spaced frequency comb [Fig.~\ref{fig:fig3}(a)]. This frequency comb is qualitatively similar to the modelocked normal GVD comb reported in~\cite{xue_normal_15}. It has several distinct features, such as a dip adjacent to the pump and asymmetrically located secondary spectral peaks with respect to the pump. In addition, complementary RF spectral measurements show that this comb exhibits low amplitude noise. The simulation [Fig.~\ref{fig:fig3}(b)] reveals that it corresponds to a stable pulse-like structure in the time domain (Fig.~\ref{fig:fig4}) with a decaying oscillatory radiation tail on each side of the intensity peak. This time domain structure also undergoes a constant drift with respect to the time reference of the LLE Eq.~(\ref{eq:LLE}). Although different mechanisms are at play, the dynamics strongly resembles that of dispersive wave emission under the influence of high-order dispersion inside resonators~\cite{milian_soliton_14,jang_observation_14,zhou_stability_15}. We confirm this hypothesis by applying to our current case the phase-matching condition which has shown to accurately predict the features of dispersive waves emitted by cavity solitons in anomalous-GVD resonators~\cite{milian_soliton_14,jang_observation_14}. Specifically, we utilize the phase-matching condition as derived in~\cite{jang_observation_14} which uses the same variable definitions and conventions as those used in Eq.~(\ref{eq:LLE}).
\begin{figure}[b]
\centerline{\includegraphics[width=0.8\columnwidth,clip]{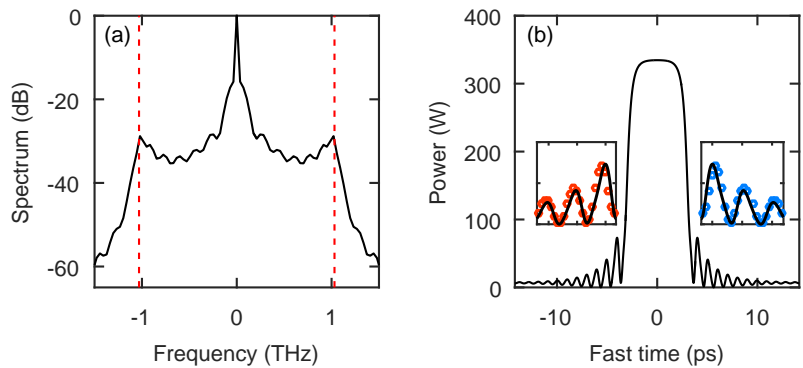}}
\caption{(a) The numerical optical spectral and (b) temporal profiles of an example of platicons numerically studied in~\cite{lobanov_frequency_15}. The insets of (b) display the close-ups of the decaying oscillatory tails which have been fitted (red and blue circles) based on our analysis.}
\label{fig:fig5}
\end{figure}

The phase-matching condition yields two complex frequencies whose real and imaginary parts describe, respectively the asymptotic frequencies and decay rates of the two oscillatory tails. This phase-matching condition includes a term that describes the drift rate of the structure which in our case is estimated from the simulation. As seen in Fig.~\ref{fig:fig3}(b), the real parts of these complex frequencies predict the positions of the secondary spectral peaks remarkably well. In addition, using the amplitudes of the oscillatory tails as fit parameters along with the derived complex frequencies, we have fitted the asymptotic power response away from the center of the pulse, again showing excellent agreement with the simulation results (red and blue circles in Fig.~\ref{fig:fig4}). Although included for completeness, it is confirmed that high-order dispersion makes a negligible contribution to the dynamics and the phase-matching condition. These results provide concrete evidence that mode coupling can manifest itself as effective high-order dispersion that breaks symmetry, leading to qualitatively similar features to those that were originally attributed to dispersive wave emission in anomalous-GVD resonators and demonstrating that the analogy between high-order dispersion and mode coupling can be generalized to the normal GVD regime.

In addition, we apply our results to several past studies where comb formation was explored under similar conditions. For specificity, we examine simulated comb-like structures referred to as platicons~\cite{lobanov_frequency_15}. Using parameters presented in~\cite{lobanov_frequency_15}, we have numerically reproduced a platicon whose optical spectral and temporal power profiles are displayed in Figs.~\ref{fig:fig5}(a) and (b) respectively. As seen in Fig.~\ref{fig:fig5}, our analysis allows accurate predictions of all the relevant features. This is expected since both our combs and platicons are produced due to mode-coupling-induced shift of the pumped resonance mode, and hence their formation can be attributed to the same underlying mechanism. The only difference is that in~\cite{lobanov_frequency_15}, only the pumped resonance was arbitrarily shifted while our analysis and results are based on the realistic scenario where the effect of mode coupling is asymmetric and occurs over several adjacent resonance modes, consistent with theoretical predictions~\cite{herr_mode_14} and experimental observations~\cite{carmon_static_08,ramelow_strong_13} of mode coupling. Our system thus provides an ideal testbed where the physics of mode-coupling-triggered comb generation can be examined under more general and realistic conditions. Note that the structure of Fig.~\ref{fig:fig5}(b) has much higher intracavity power than our result since the pump power is higher and the platform under consideration is a crystalline microresonator which is known to have higher quality factors in comparison to on-chip $\mathrm{Si}_3\mathrm{N}_4$ ring resonators. Finally, it should be pointed out that our system may also have connections to dark pulse structures~\cite{parra_origin_16} whose origin and existence were explained in terms of interlocked switching waves~\cite{rozanov_transverse_82,coen_convection_99}. These switching waves exhibit decaying oscillatory asymptotic profiles like those shown in Figs.~\ref{fig:fig4} and~\ref{fig:fig5}(b), and can hence be analyzed by similar means~\cite{parra_origin_16}. Such a connection suggests an alternative interpretation of the temporal profile shown in Fig.~\ref{fig:fig4} consisting of a dark structure with a broad width.

\section{Broadband microresonator comb generation in normal GVD}

\begin{figure}[t]
\centerline{\includegraphics[width=0.8\columnwidth,clip]{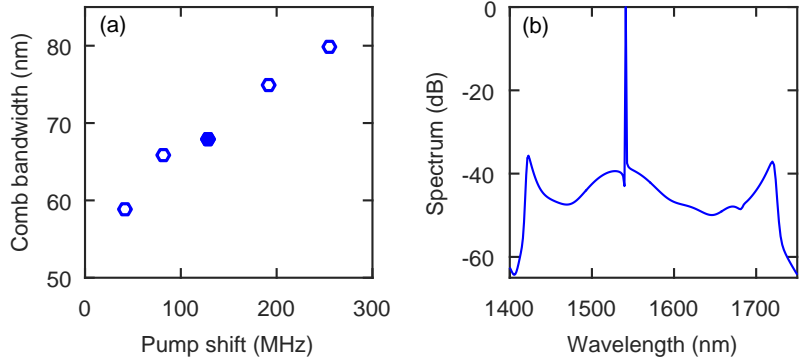}}
\caption{(a) Numerically estimated correlation between the coupling strength and the comb bandwidth. The solid circle corresponds to our experimental case. (b) Simulated broadband comb with a bandwidth of over $300$~nm.}
\label{fig:fig6}
\end{figure}

We numerically explore the possibility of generating broadband normal GVD comb under similar conditions. Increasing the pump power to 50~mW while keeping all other parameters the same, we first investigate correlation between the coupling strength and the bandwidth of the resulting comb. The result is summarized in Fig.~\ref{fig:fig6}(a), and there is clearly an approximately linear relationship. This trend can be understood as follows. Under the influence of purely normal GVD, the resonator FSR decreases monotonically with the optical frequency and resonances progressively deviate from their equidistant positions (defined in the absence of GVD). The further we move away from the pump, the larger the deviations. As a consequence, in order to phase-match the MI process involving modes that are further away from the pump, the pump mode itself must be shifted by a larger amount, such that equivalent anomalous GVD is achieved among the pump and the sideband modes~\cite{savchenkov_kerr_12,xue_modelocked_15}. Another important issue we must consider is the GVD, as it is known that the comb bandwidth follows a scaling law in the form of $\propto 1/\sqrt{|\beta_2|}$~\cite{coen_universal_13}, i.e. the second-order GVD parameter $|\beta_2|$ must be small. We therefore reduce our $\beta_2$ to $1.8~\mathrm{ps^2/km}$ and double the coupling strength which corresponds to the last point in Fig.~\ref{fig:fig6}(a), still an experimentally feasible value (shift of $\simeq 260$~MHz). These conditions enable generation of a broadband comb spanning over 300~nm, as plotted in Fig.~\ref{fig:fig6}(b), which is comparable to dark-pulse frequency combs examined in~\cite{xue_modelocked_15}. Note that efficiency (comb-to-pump power ratio) appears low since this comb corresponds to the output of the resonator at the through-port. This issue can be circumvented by operating in the over-coupled regime~\cite{delhaye_optical_07} or by extracting the comb from a drop port~\cite{wang_intracavity_16}. Filtering and amplification followed by external broadening may lead to even broader combs.

\section{Conclusion}
In conclusion, we examine the dynamics of comb operation in the normal GVD regime enabled by mode coupling in a resonator under several scenarios. We explore a previously unidentified regime outside the Kerr-bistability where mode coupling triggers intracavity MI and also propose a modified phase-matching condition that describes the features of the MI spectrum. It is demonstrated that mode coupling can overcome the MI power threshold requirement. We also observe and theoretically analyze the transition of the intracavity light from the MI spectrum to a low-noise comb, establishing connections between our results and several past studies carried out under similar conditions. Our results therefore provide invaluable insight into the dynamics that underlie microresonator comb generation influenced by mode coupling under normal GVD conditions.


\section*{Acknowledgements}
The authors gratefully acknowledge support from the Air Force Office of Scientific Research under award number FFA9550-15-1-0303 and the Advanced Research Projects Agency-Energy (DE-AR0000720). This work was performed in part at the Cornell Nanoscale Facility, a member of the National Nanotechnology Infrastructure Network, which is supported by the NSF (grant ECS-0335765).

\end{document}